\documentclass[aps,prb,twocolumn]{revtex4}

\usepackage{amssymb}
\usepackage{graphicx}
\usepackage{hyperref}
\begin{document}

\title{Strong coupling superconductivity due to massless boson exchange }
\author{Andrey V. Chubukov$^{(1)}$ and J\"{o}rg Schmalian$^{(2)}$}
\affiliation{$^{(1)}$Department of Physics and Condensed Matter Theory Center, University
of Maryland, College Park, MD 20742\\
$^{(2)}$Department of Physics and Astronomy and Ames Laboratory, Iowa State
University, Ames, IA 50011}
\date{\today }

\begin{abstract}
We solve the problem of fermionic pairing mediated by a massless boson in
the limit of large coupling constant. At weak coupling,
the transition temperature is exponentially small  and superconductivity 
is robust against phase fluctuation. In the strong coupling limit, 
the pair formation occurs at a temperature of the order of the Fermi energy,
 however, the actual transition temperature is much
smaller due to phase and amplitude fluctuations of the pairing gap.
 Our model calculations describe 
 superconductivity due to color magnetic interactions 
in quark matter and in systems close to a ferromagnetic quantum critical
 point with Ising symmetry. Our  strong coupling results are, however,
more general and can be applied to other systems as well, including the
antiferromagnetic exchange in 2D used for description of the cuprates.  
\end{abstract}

\pacs{71.10Li, 74.20Mn}
\maketitle




\section{Introduction}

\label{sec:introduction}

Strong coupling superconductivity due to the interaction between electrons
and lattice vibrations has been \ successfully studied using the coupled
Eliashberg equations\cite{Eliashberg60} for the frequency dependent normal
and anomalous self energies of superconductors\cite%
{Scalapino69,McMillan68,Marsiglio03}. The theory finds its justification in
the weakness of the corrections to the electron-phonon vertex, caused by the
small ratio of the electron and ion masses\cite{Migal58}. This theory
inspired numerous efforts to describe superconductivity caused by other
bosons\cite%
{Emery64,Berk66,Anderson,Anderson2,Leggett75,Emery86,Miyake86,Scalapino86,Monthoux91,Scalapino95,pines97}%
, even though its justification turns out to be considerably more subtle in
some of those cases\cite{Schrieffer95,Chubukov95,Altshuler95}. Important
progress has been made in the study of pairing due to the exchange of bosons
that are collective excitations of the fermions\cite%
{Emery64,Berk66,Anderson,Anderson2,Leggett75,Emery86,Miyake86,Scalapino86,Monthoux91,Scalapino95,pines97}%
. In this context, the interplay between superconductivity and quantum
criticality is particularly interesting\cite{Abanov01,Roussev01,Wang01,Dzero04,Abanov03,Chubukov03} as
superconductivity in correlated electron systems often occurs in the
proximity of a quantum critical point (QCP) \cite%
{Mathur98,Maeno94,Saxena98,Aoki01,Fisk98}. At a QCP, the pairing boson
becomes massless, and new and unexpected behavior emerges\cite%
{Abanov01,Son99,Chubukov03}. A related problem occurs in the theory of
quantum chromodynamics at high density where single-gluon exchange becomes
dominant\cite{Gross73}. The exchange of gluons is believed to cause color
superconductivity\cite{Bailin84,Alford98,Rapp98}. As was pointed out by Son%
\cite{Son99}, and later in Refs.\cite{Schaefer99,Pisarski00}, the color
magnetic interaction in high density QCD is unscreened at low temperatures,
i.e. the pairing is mediated by a gapless boson. The pairing problem then
becomes formally very similar to superconductivity at a QCP, even though the
transition temperatures may be different by a factor as big as $10^{12}$.

In previous studies of the pairing problem near a QCP, the authors of Refs. 
\cite{Abanov01,Son99,Chubukov03} assumed that the effective, boson-mediated
fermion-fermion interaction {$u$} is much smaller than the fermionic
bandwidth, $W$ (which is generally of the same order as $E_{F}$), i.e., 
\begin{equation}
g\sim \frac{u}{W}\ll 1.  \label{e_1}
\end{equation}%
The limit $g\ll 1$ is often called weak coupling. This notation is not quite
correct, as near a QCP the smallness of $g$ doesn't imply that the system
behaves as a weakly coupled Fermi liquid -- the mass renormalization due to
the exchange of a gapless boson is still singular in $D\leq 3$ and destroys
the Fermi liquid behavior at the QCP (see below). To simplify the notations,
we nevertheless refer to $g\ll 1$ as weak coupling and $g\gg 1$ as strong
coupling. With this notation the Eliashberg theory for superconductivity due
to electron-phonon interaction\cite{Scalapino69,McMillan68,Marsiglio03} is
in the "weak coupling" limit since $g\propto \frac{v_{s}}{v_{F}}\lambda $
 is small. This is due to the smallness of the ratio 
$\frac{v_{s}}{v_{F}}$ of the sound and Fermi velocities, while the product $\lambda
=\rho _{F}V_{\mathrm{ep}}$ of the electron-phonon interaction $V_{\mathrm{ep}%
}$ and the density of states at the Fermi level $\rho_{F}$ can be of order
unity.

The condition $g\ll 1$ implies that the pairing comes only from fermions in
a tiny range of momenta around the Fermi surface, i.e., that the system
behavior at energies comparable to $E_{F}$ is irrelevant for the pairing.
This makes the pairing problem universal and allows one to use
well-established computational techniques, e.g. the Eliashberg theory\cite%
{Eliashberg60}. However, for various systems of interest the interaction is
not necessarily small. In particular, the same interaction that leads to the
pairing is often also responsible for the onset of order at the QCP. Generic
density-wave instabilities come from fermions with energies $O(E_{F})$ and
require $g$ to be of order unity\cite{comm_1}. In the cuprate
superconductors, to which the ideas of collective-mode mediated $d-$wave
pairing was applied\cite{Scalapino95,pines97,Abanov03}, the Hubbard
interaction $U$ is at least comparable to $W$ as is evidenced by e.g. the
Heisenberg antiferromagnetism at half-filling. For color superconductivity,
the effective coupling {$u$} is also not necessarily small compared to $%
E_{F} $, and $g$ well may be larger than 1.

\begin{figure}[tbp]
\label{Fig1} \includegraphics[width=8.5cm]{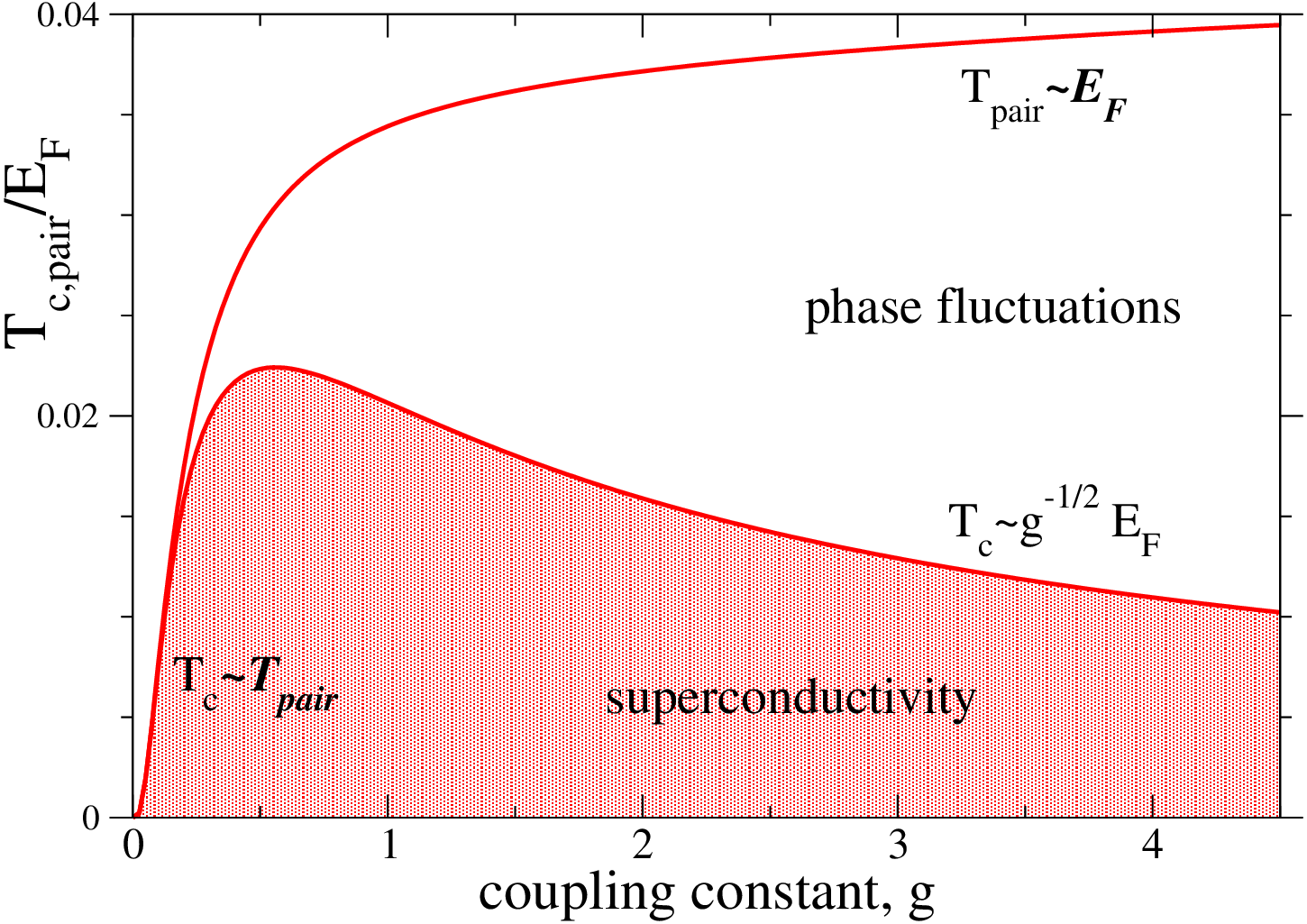}
\caption{Schematic phase diagram for the superconducting transition
temperature $T_{c}$ and the pairing instability temperature $T_{\mathrm{pair}%
}$ as function of the dimensionless coupling constant $g$. While $%
T_{c}\simeq T_{\mathrm{pair}}$ for weak coupling, an intermediate regime $%
T_{c}<T<T_{\mathrm{pair}}$ with phase (and amplitude) fluctuations occurs
 in the strong coupling limit.}
\end{figure}

These arguments call for an understanding of the pairing problem beyond the
\textquotedblleft weak coupling\textquotedblright\ limit. In the present
paper we extend previous \textquotedblleft weak-coupling\textquotedblright\
studies of the pairing mediated by a gapless boson to the truly strong
coupling limit $g\gg 1$. For definiteness we consider pairing of 3D
electrons mediated by a scalar boson which is gapless at $q=0$. This model
describes $p$-wave superconductivity near a ferromagnetic Ising QCP, and
color superconductivity of quarks. However, the results at strong coupling
are quite general and can be applied to other systems as well, including the
antiferromagnetic exchange in 2D used for description of the cuprates. We
discuss applications to other systems in a separate section. There is also a
connection between our model and the interaction between conduction
electrons, mediated by transverse photons~ \cite{Holstein73,Reizer89}.
However, as shown in \cite{Reizer89},\ the exchange of transverse photons
does not lead to superconductivity.

A word of caution. In the context of quark pairing mediated by gluons, the
equation for the pairing vertex has only been derived in a gauge invariant
manner at weak coupling.\cite{Pisarski02}. At strong coupling, antiparticle
pairing, neglected in our model, may come into play. Still, qualitatively,
the results obtained assuming only particle pairing likely remain valid at
both weak and strong coupling.

The main results of this paper are summarized in the phase diagram, Fig.\ref%
{Fig1}. In the weak coupling limit, we find, in agreement with Son~\cite%
{Son99} and others~\cite{Schaefer99,Pisarski00}, that the transition
temperature behaves, to leading exponential order, as 
\begin{equation}
\log \frac{\omega _{0}}{T_{c}}=\frac{\pi }{2\sqrt{g}},  \label{son1}
\end{equation}%
where $\omega _{0}\sim E_{F}/g$. This result is parametrically larger than
the usual BCS result\cite{BCS57}: $\log {\omega _{0}}/{T_{c}}\propto 1/g$,
and the difference is due to the gapless nature of the pairing boson (see
below). Still, $T_{c}$ is exponentially small at small $g$.

At $g=O(1)$, $T_{c}$ becomes of order $E_{F}$, although the prefactor for $%
T_{c}$ is a small number. At even larger $g$, we find two characteristic
temperatures. The larger temperature, $T_{\mathrm{pair}}$, sets the onset of
pairing, and is of order $E_{F}$ (again, with a small prefactor). The
smaller temperature, $T_{c}$ is of order $\sqrt{\omega _{0}E_{F}}\sim E_{F}/%
\sqrt{g}\ll E_{F}$. This temperature is determined by the superfluid
stiffness, and sets the scale for phase coherence, i.e., of the actual
superconductivity. In between $T_{c}\sim E_{F}/\sqrt{g}$ and $T_{\mathrm{pair%
}}\sim E_{F}$, the system displays pseudogap behavior: pairs of fermions are
already formed, but do not move coherently.

The accuracy of the computations is a central issue for the theoretical
analysis near a QCP. At small $g$, one can use Eliashberg theory since the
relevant bosonic and fermionic frequencies are much smaller than $E_{F}$.
Although the frequency-dependent self-energy is not small due to the
near-criticality, vertex corrections and the momentum-dependent self-energy
are exponentially small (see below). At $g=O(1)$, typical frequencies become
of order $E_{F}$, vertex corrections become $O(1)$, and the
momentum-dependent self-energy becomes of the same order as the
frequency-dependent self-energy. In this situation, no reliable theoretical
scheme is possible.

A naive expectation would be that at strong coupling, vertex corrections get
even stronger. We show, however, that at larger $g$, vertex corrections
actually saturate at a value $O(1)$ and do not grow with $g$. At the same
time, the momentum dependent term in the self-energy again becomes small
compared to its frequency dependence, this time the relative smallness is in 
$1/g$. Furthermore, the pairing problem at $g\gg 1$ still involves fermions
with energies below $E_{F}$ for which the density of states can be
approximated by a constant. As a result, the new version of the Eliashberg
theory (more accurately, the local theory) becomes qualitatively valid. This
new local theory is different from the original Eliashberg theory in that
the lattice cannot be neglected, and the existence of a finite bosonic
bandwidth now plays a crucial role. Still, like in the Eliashberg theory, we
derive closed-form equations for the fermionic self-energy and the pairing
vertex. To justify the local approximation at $g>1$ quantitatively, we
extend the model to $N$ fermion flavors and consider the limit of large $N$%
\cite{Abanov03}. In this case, vertex corrections become of order $1/N$ and
can be safely neglected.

The structure of the paper is as follows. In the next section we set up the
model and define the large $N$ limit used to perform the strong coupling
calculation. In Sec. III we briefly discuss the weak coupling limit and
present an alternative derivation of Son's result\cite{Son99} for $T_{c}$.
In Sec. IV we solve the pairing problem at strong coupling. In Sec. V we
analyze gap fluctuations and demonstrate the existence of two characteristic
\ temperature scales. In Sec VI we justify our computational procedure. In
Sec VII we discuss other systems, including the cuprates. The last section
presents our conclusions. Several technical details are presented in the
Appendix.

\section{Model and large $N$ expansion}

We consider the pairing problem in which 3D fermions, $\psi _{k}$, interact
via exchanging a massless, bosonic mode with a static propagator $%
D_{q}^{\left( 0\right) }=1/q^{2}$: 
\begin{equation}
\mathcal{H}_{\mathrm{int}}=-\frac{{u}}{k_{F}}~\sum_{k,k^{\prime },q}\psi
_{k+q}^{\dagger }\psi _{k^{\prime }-q}^{\dagger }D_{q}^{\left( 0\right)
}\psi _{k^{\prime }}\psi _{k}.  \label{Hint2}
\end{equation}%
Here {$u$}$>0$ is the effective interaction (with the dimension of energy),
and $k_{F}$ is the Fermi momentum. Note the overall sign in (\ref{Hint2}) is
opposite to that in systems with Coulomb interaction~\cite{yakov}. Another
energy scale in the problem is the fermionic bandwidth $W$ (roughly, the
scale up to which the fermionic dispersion $\varepsilon _{k}$ can be
linearized around Fermi surface, $\varepsilon _{k}=v_{F}(k-k_{F})$). The
ratio of the two characteristic energies defines the dimensionless coupling
constant $g$ in Eq. (\ref{e_1}).

The interaction (\ref{Hint2}) leads to pairing, and also gives rise to
fermionic and bosonic self-energies, $\Sigma _{k}(\omega )$ and $\Pi
_{q}(\Omega )$, respectively. The two self-energies are related to fermionic
and bosonic propagators via 
\begin{eqnarray}
G_{k}^{-1}(\omega ) &=&i\omega -v_{F}(k-k_{F})-\Sigma _{k}(\omega ) 
\nonumber \\
D_{q}^{-1}(\Omega ) &=&q^{2}+\Pi _{q}\left( \Omega \right) .  \label{e_3}
\end{eqnarray}

In order to perform a controlled calculation at strong coupling, we
generalize the model of Eq. (\ref{e_3}) to $N$ fermion flavors and rescale $%
v_{F}\rightarrow v_{F}N$ and $u\rightarrow ${$u$}$N$. In what follows, we
assume that the new $v_{F}$ and {$u$} are constants, independent on $N$.\cite%
{Abanov03}

The generalized Eliashberg theory is a set of three coupled integral
equations for the pairing vertex $\Phi $ and the self-energies $\Sigma $ and 
$\Pi $. We will primarily be interested in the onset of the pairing and
consider the linearized equation for $\Phi $, and normal state expressions
for $\Sigma $ and $\Pi $. Then the system of three coupled equations has the
form 
\begin{eqnarray}
\Phi \left( \omega \right) &=&\frac{u}{k_{F}}\int_{q,\omega ^{\prime
}}D_{q}\left( \omega -\omega ^{\prime }\right) \Phi \left( \omega ^{\prime
}\right)  \nonumber \\
&&\times G_{k_{F}+q}\left( \omega ^{\prime }\right) G_{k_{F}+q}\left(
-\omega ^{\prime }\right) ,  \nonumber \\
\Sigma \left( \omega \right) &=&\frac{u}{k_{F}}\int_{q,\omega ^{\prime
}}D_{q}\left( \omega -\omega ^{\prime }\right) G_{k_{F}+q}(\omega ^{\prime
}),  \nonumber \\
\Pi _{q}\left( \Omega \right) &=&\frac{u}{k_{F}}\int_{k,\omega }G_{k}(\omega
)G_{k+q}(\omega +\Omega ).  \label{ELmom}
\end{eqnarray}%
We used the notation $\int_{q,\omega }...=\int \frac{d^{3}q}{\left( 2\pi
\right) ^{3}}T\sum_{\omega _{n}}...$ with Matsubara frequencies $\Omega
_{n}=2n\pi T$ \ and $\omega _{n}=\left( 2n+1\right) \pi T$ for bosons and
fermions, respectively.

Eqs. (\ref{ELmom}) neglect vertex corrections and the momentum dependence of 
$\Sigma $ and $\Phi $. We will argue below that at large $N$, both
approximations hold both at weak and at strong coupling. Physically, these
approximations are based on the (verifiable) assumption that bosons are slow
modes compared to fermions. This allows one to factorize the momentum
integration in Eqs.(\ref{ELmom}). Namely, for every given $\mathbf{k}_{F}$
along the Fermi surface, the integration over the component $q_{\perp }$
transverse to Fermi surface in the equations for $\Phi $ and $\Sigma $
involves only fast fermions, while integrating over the remaining two
momentum components $q_{\parallel }$ in the bosonic propagator, one can set $%
q_{\perp }=0$. This implies that the boson propagator actually only appears
in Eqs. (\ref{ELmom}) through the "local"\ interaction 
\begin{equation}
d\left( \Omega \right) =\int_{0}^{q_{0}}q_{\parallel }dq_{\parallel }D_{%
\mathbf{q}_{\parallel },q_{\perp }=0}(\Omega ),  \label{may_29_4}
\end{equation}%
where $q_{0}\sim k_{F}$ is the upper cutoff in the integral over $%
q_{\parallel }$. As a result, the equations for $\Phi $ and $\Sigma $ in (%
\ref{ELmom}) reduce to 
\begin{eqnarray}
\Phi \left( \omega \right) &=&\frac{3g}{2}\int d\omega ^{\prime }\frac{\Phi
\left( \omega ^{\prime }\right) d\left( \omega -\omega ^{\prime }\right) }{%
\left\vert \omega ^{\prime }+i\Sigma \left( \omega ^{\prime }\right)
\right\vert }  \nonumber \\
\ \Sigma \left( \omega \right) &=&-i\frac{3g}{2}\int d\Omega \mathrm{sign}%
\left( \Omega +\omega \right) d\left( \Omega \right) ,  \label{ELmom_1}
\end{eqnarray}%
where the factor of $\frac{3}{2}$ is for further convenience and the coupling
constant $g$ is given as%
\begin{equation}
g=\frac{u}{24\pi ^{2}E_{F}^{\ast }},  \label{e_5}
\end{equation}%
with $E_{F}^{\ast }=v_{F}k_{F}/2$. Without further approximation we can
explicitly solve for $\Pi \left( \Omega \right) $ and find\cite{Abanov03}%
\begin{equation}
\Pi (q,\Omega )=\gamma \frac{|\Omega |}{q},  \label{bosself}
\end{equation}%
where 
\begin{equation}
\gamma =12\pi ^{2}k_{F}^{3}~\frac{g}{E_{F}^{\ast }}.  \label{e_5_1}
\end{equation}%
Note that $\gamma $ does not depend on $N$, despite the fact that the Landau
damping term contains a flavor index $N$ as an overall factor. The $N$%
-independence of $\gamma $ is the result of our rescaling: in rescaled
variables $\gamma \rightarrow N\frac{uNk_{\mathrm{F}}}{\left( v_{F}N\right)
^{2}}$ stays finite.

For the "local" \ interaction (\ref{may_29_4}), we obtain from Eqs. (\ref%
{e_3}) and (\ref{bosself}): 
\begin{equation}
d\left( \Omega \right) =\frac{1}{3}\log \left( 1+\frac{\omega _{0}}{%
\left\vert \Omega \right\vert }\right) ,  \label{kernel}
\end{equation}%
where $\omega _{0}$ is the characteristic frequency of the bosonic degrees
of freedom: 
\begin{equation}
\omega _{0}=\frac{q_{0}^{3}}{\gamma }=\frac{E_{F}}{g}  \label{e_6}
\end{equation}%
and we introduced 
\begin{equation}
E_{F}=\frac{E_{F}^{\ast }}{12}~\left( \frac{q_{0}}{\pi k_{F}}\right) ^{3}.
\label{e_6_1}
\end{equation}%
Below we will refer to $E_{F}$ as to Fermi energy. We should keep in mind
however that our $E_{F}$ depends on the choice of the upper momentum cut off 
$q_{0}$ and is only of the same order of magnitude as a actual Fermi energy
of the system.

\begin{figure}[tbp]
\label{Fig2} \includegraphics[width=6.5cm]{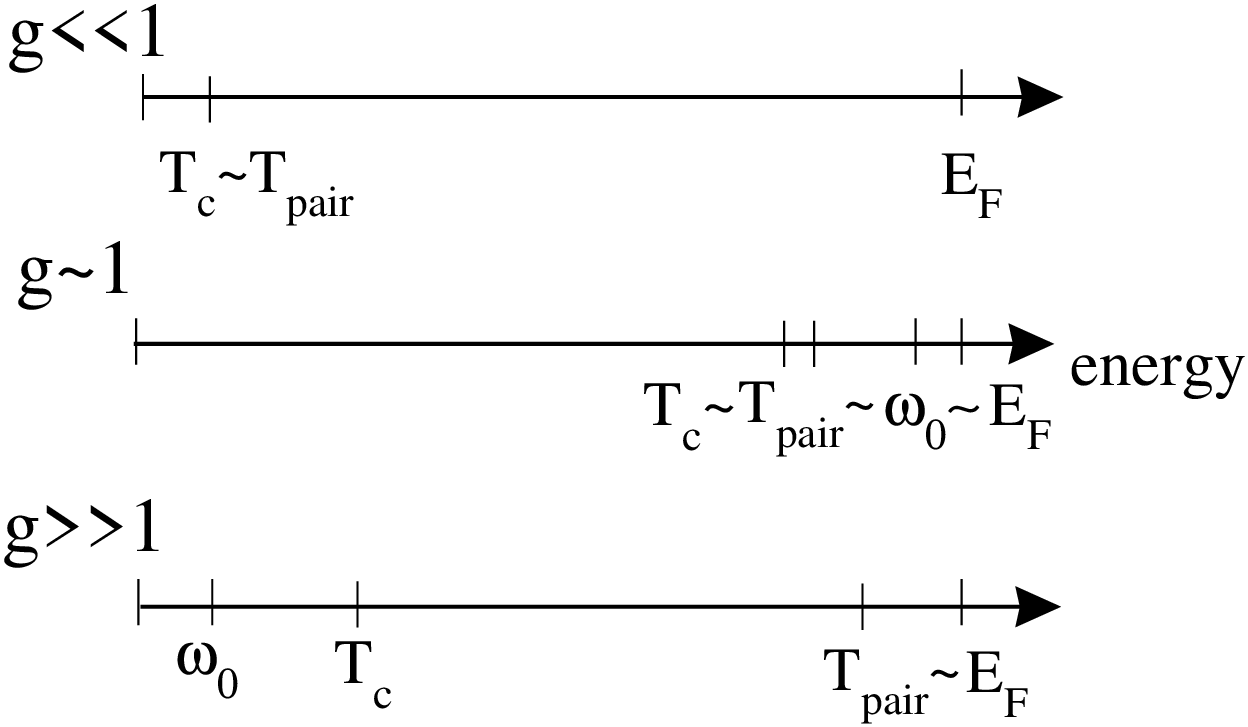}
\caption{ Characteristic energy scales for the weak ($g\ll 1$), intermediate
($g\sim 1$) and strong \ ($g\gg 1$) coupling limit. While at weak coupling, $%
\protect\omega _{0}\simeq E_{F}/g$ is large compared to $E_{F}$ and thus
irrelevant, it emerges as a new low energy scale in the strong coupling
limit. }
\end{figure}

Substituting Eq.(\ref{kernel}) into Eq.(\ref{ELmom}) and integrating over
frequency, we obtain 
\begin{eqnarray}
i\Sigma \left( \omega \right) &=&\omega g\left( \frac{\omega _{0}}{%
\left\vert \omega \right\vert }\log \frac{\omega _{0}+\left\vert \omega
\right\vert }{\omega _{0}}+\log \frac{\omega _{0}+\left\vert \omega
\right\vert }{\left\vert \omega \right\vert }\right)  \nonumber \\
&=&\left\{ 
\begin{array}{cc}
\omega g\log \frac{\omega _{0}}{\left\vert \omega \right\vert }\  & 
\left\vert \omega \right\vert \ll \omega _{0} \\ 
\mathrm{sign}\left( \omega \right) g\omega _{0}\log \frac{\left\vert \omega
\right\vert }{\omega _{0}} & \left\vert \omega \right\vert \gg \omega _{0}%
\end{array}%
\right. .  \label{e_7}
\end{eqnarray}%
We see that $\omega _{0}$ sets the scale at which the momentum cutoff in the
bosonic propagator begins affecting the fermionic self-energy. At low
energies, the cutoff is irrelevant, and the self-energy has the form typical
for a marginal Fermi liquid\cite{Varma90}. At $\omega >\omega _{0}$, the
self energy almost saturates and only logarithmically depends on frequency.
At weak coupling, $\omega _{0}=E_{F}/g>E_{F}$, and the crossover is
meaningless as Eq. (\ref{e_7}) only holds up to $\omega \sim E_{F}$ (we
recall that in obtaining Eq.(\ref{e_7}) we approximated the density of
states by a constant). The marginal Fermi liquid behavior then extends all
the way up to $E_{F}$. At strong coupling $\omega _{0}\ll E_{F}$, and the
crossover in $\ \Sigma (\omega )$ occurs well below $E_{F}$. In this
situation, marginal Fermi liquid behavior only holds at small frequencies $%
\omega <E_{F}/g$, while at $E_{F}/g<\omega <E_{F}$, $\Sigma (\omega )$
depends logarithmically on frequency (see Fig. \ref{Fig2}).

The crossover in the self-energy at strong coupling parallels the crossover
in the \textquotedblleft local\textquotedblright\ bosonic propagator $%
d(\omega )$ in Eq.(\ref{kernel}) 
\begin{equation}
d\left( \Omega \right) =\left\{ 
\begin{array}{cc}
\frac{1}{3}\log \frac{\omega _{0}}{\left\vert \Omega \right\vert } & 
\left\vert \Omega \right\vert \ll \omega _{0} \\ 
\frac{1}{3}~\frac{\omega _{0}}{\left\vert \Omega \right\vert } & \left\vert
\Omega \right\vert \gg \omega _{0}%
\end{array}%
\right. .  \label{kernel_1}
\end{equation}
Like for the self-energy, this crossover is meaningful only at strong
coupling, when $\omega_0 < E_F$.

Substituting the self-energy and $d(\Omega )$ into the equation for the
pairing vertex, we obtain 
\begin{equation}
\Phi \left( \omega \right) =\frac{g}{2}\int d\omega ^{\prime }\frac{\Phi
\left( \omega ^{\prime }\right) }{|\omega ^{\prime }+i\Sigma (\omega
^{\prime })|}\log \left( 1+\frac{\omega _{0}}{\left\vert \omega -\omega
^{\prime }\right\vert }\right) .  \label{e_8}
\end{equation}%
Strictly speaking, we have to evaluate this equation at finite $T$, because
the linearized equation for $\Phi $ is only valid at the onset temperature
for the pairing. By reasons that we outline below, we label this temperature
as $T_{\mathrm{pair}}$ rather than $T_{c}$. As we will only be interested in
the order of magnitude estimate for $T_{\mathrm{pair}}$, we adopt a
simplified approach, and instead of performing the discrete Matsubara sum,
use Eq.(\ref{e_8}) at finite $T$, but introduce a lower frequency cutoff at $%
\omega \sim T$. In the weak coupling limit, this procedure was shown earlier~%
\cite{Pisarski00} to yield the same $T_{\mathrm{pair}}$ (modulo a numerical
prefactor), as one would obtain by performing an explicit summation over
discrete Matsubara frequencies. In Appendix B we show that the same holds
for large $g$. With this simplification we have to solve 
\begin{equation}
\Phi \left( \omega \right) =g\int_{T_{\mathrm{pair}}}^{\infty }d\omega
^{\prime }\frac{\Phi \left( \omega ^{\prime }\right) }{|\omega ^{\prime
}+i\Sigma (\omega ^{\prime })|}~K\left( \omega ,\omega ^{\prime }\right) ,
\label{elcomp}
\end{equation}%
with bosonic kernel 
\begin{equation}
K\left( \omega ,\omega ^{\prime }\right) =\frac{1}{2}\log \left[ \left( 1+%
\frac{\omega _{0}}{\left\vert \omega -\omega ^{\prime }\right\vert }\right)
\left( 1+\frac{\omega _{0}}{\left\vert \omega +\omega ^{\prime }\right\vert }%
\right) \right] .  \label{dwwp}
\end{equation}%
In what follows we solve this equation, first in the weak coupling limit $%
g\ll 1$, where we reproduce the results of Ref.\cite{Son99}, and then in the
strong coupling limit $g\gg 1$.

\section{Pairing problem at weak coupling}

At weak coupling, one obviously expects $T_{\mathrm{pair}}$ to be much
smaller than $\omega _{0}$ (see Fig.\ref{Fig2}). This in turn implies that
only frequencies $\omega \ll \omega _{0}$ are relevant. For these
frequencies, the self-energy $\Sigma (\omega )$ and the kernel $K\left(
\omega ,\omega ^{\prime }\right) $ in (\ref{dwwp}) can be simplified to 
\begin{eqnarray}
\Sigma \left( \omega \right) &=&-i\omega g\log \frac{\omega _{0}}{\left\vert
\omega \right\vert }  \nonumber \\
K\left( \omega ,\omega ^{\prime }\right) &=&\log \frac{\omega _{0}}{\sqrt{%
\left\vert \omega ^{2}-\omega ^{\prime }{}^{2}\right\vert }}.  \label{dwwp_1}
\end{eqnarray}%
Eq. (\ref{elcomp}) then becomes 
\begin{equation}
\Phi \left( \omega \right) =g\int_{T_{\mathrm{pair}}}^{\omega _{0}}d\omega
^{\prime }\frac{\Phi \left( \omega ^{\prime }\right) \log \frac{\omega _{0}}{%
\sqrt{\left\vert \omega ^{2}-\omega ^{\prime }{}^{2}\right\vert }}}{\omega
^{\prime }(1+g\omega ^{\prime }~\log \frac{\omega _{0}}{\omega ^{\prime }})}.
\label{elcomp_1}
\end{equation}%
This equation yields $T_{\mathrm{pair}}$ for the pairing in a marginal Fermi
liquid. Eq. (\ref{elcomp_1}) was solved numerically in Ref.\cite{varma}. We
show that an analytic solution is also possible. Our computational procedure
is similar to the one used by Son~\cite{Son99}. In addition to the approach
of Ref.\cite{Son99} we also analyze the pairing susceptibility.

If the two logarithmic terms in the r.h.s. of Eq.(\ref{elcomp_1}) were
absent, the equation for the pairing vertex would be the same as in BCS
theory\cite{BCS57}, and $T_{\mathrm{pair}}$ would scale as $\omega
_{0}e^{-1/g}$. However, as Son demonstrated~\cite{Son99}, the presence of
the logarithm in the pairing kernel substantially enhances $T_{\mathrm{pair}%
} $ at weak coupling and changes its functional form to $T_{\mathrm{pair}%
}\propto \omega _{0}e^{-\pi /(2\sqrt{g})}$. The easiest way to see this is
to introduce logarithmic variables: $x=\log \left( \frac{\omega _{0}}{\omega 
}\right) $, $x^{\prime }=\log \left( \frac{\omega _{0}}{\omega ^{\prime }}%
\right) $, $x_{T}=\log \left( \frac{\omega _{0}}{T}\right) $, and re-write
Eq. (\ref{elcomp_1}) with logarithmic accuracy as 
\begin{equation}
\Phi (x)=g\int_{0}^{x}dx^{\prime }\frac{x^{\prime }\Phi \left( x^{\prime
}\right) }{1+gx^{\prime }}+gx\int_{x}^{x_{T}}dx^{\prime }\frac{\Phi \left(
x^{\prime }\right) }{1+gx^{\prime }}.  \label{elcomp_2}
\end{equation}%
Differentiating both sides of Eq. (\ref{elcomp_2}) over $x$, we find 
\begin{equation}
\frac{d\Phi \left( x\right) }{dx}=g\int_{x}^{x_{T}}\frac{\Phi \left(
x\right) }{1+gx}.  \label{diffeq_a}
\end{equation}%
Differentiating one more time, we find that the integral equation for the
anomalous vertex reduces to a second-order differential equation: 
\begin{equation}
\frac{d^{2}\Phi \left( x\right) }{dx^{2}}=-g\frac{\Phi \left( x\right) }{1+gx%
}.  \label{diffeq}
\end{equation}%
The $gx$ term in the r.h.s. of Eq. (\ref{diffeq}) is due to the fermionic
self energy. We assume and verify afterwards that $gx\ll 1$ for all relevant 
$x$, and drop this term from (\ref{diffeq}). The solution of Eq. (\ref%
{diffeq}) is then elementary: 
\begin{equation}
\Phi \left( x\right) =A\cos \left( \sqrt{g}x\right) +B\sin \left( \sqrt{g}%
x\right) .  \label{fweak}
\end{equation}%
The two boundary conditions 
\begin{eqnarray}
\Phi \left( x=0\right) &=&0,  \label{boundary} \\
\left. \frac{d\Phi (x)}{dx}\right\vert _{x=x_{T}} &=&0  \nonumber
\end{eqnarray}%
follow from (\ref{elcomp_2}) and (\ref{diffeq_a}), respectively. They yield $%
A=0$, and 
\begin{equation}
\cos (\sqrt{g}x_{T})=0.  \label{e_9}
\end{equation}%
The onset temperature $T_{\mathrm{pair}}$ corresponds to the smallest $x_{T}$
that satisfies (\ref{e_9}), i.e. to $x_{T}\sqrt{g}=\pi /2$. Ignoring
pre-exponential factors, we then reproduce Son's result\cite{Son99}: 
\begin{equation}
T_{\mathrm{pair}}\simeq \omega _{0}e^{-\frac{\pi }{2\sqrt{g}}}.  \label{wc}
\end{equation}%
The relevant value of $x$ are $x_{T}\propto 1/\sqrt{g}$, i.e., $gx\sim
gx_{T}\sim \sqrt{g}\ll 1$. This justifies dropping the $gx$ term (i.e.,
fermionic self-energy) from (\ref{diffeq}). The first order correction to $%
T_{\mathrm{pair}}$ due to the self-energy was analyzed in Ref.\cite{Wang02}
in the context of color superconductivity.

To get more insight into the pairing instability, it is also instructive to
analyze the pairing susceptibility $\chi _{pp}(\omega ,T)$. This is done by
adding an infinitesimally small external pairing field $\Phi _{0}$ to the
r.h.s. of (\ref{elcomp_1})~\cite{comm_2}. The pairing susceptibility is 
\[
\chi _{pp}(\omega ,T)=\left. \frac{\partial \Phi (\omega ,T)}{\partial \Phi
_{0}}\right\vert _{\Phi _{0}\rightarrow 0}=\frac{\Phi (\omega ,T)}{\Phi _{0}}%
. 
\]

\begin{figure}[tbp]
\label{Fig3} \includegraphics[width=8.5cm]{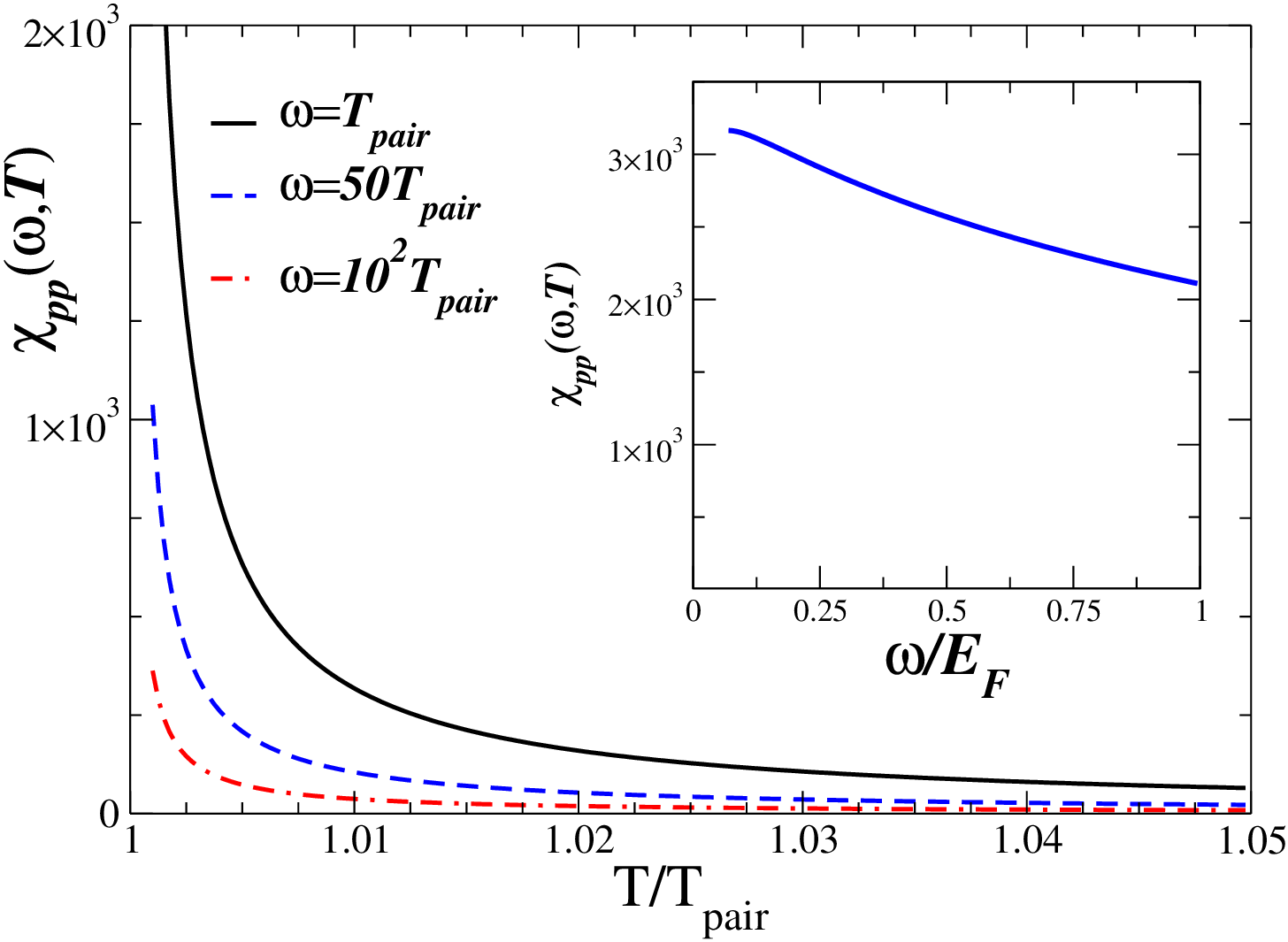}
\caption{Particle-particle response function $\protect\chi _{pp}\left( 
\protect\omega ,T\right) $ for $g=0.1$ as function of $T$ for various
energies. Insert: $\protect\chi _{pp}\left( \protect\omega ,T\right) $ as as
function of energy for $T=1.001T_{c}$. }
\end{figure}

At $T>T_{\mathrm{pair}}$, $\Phi (\omega )\propto \Phi _{0}$, and the pairing
susceptibility is finite. If the transition is of second order, $\chi
_{pp}(\omega ,T)$ diverges at $T_{\mathrm{pair}}$. In BCS theory, $\Phi
(\omega ,T)$ does not depend on frequency, and $\chi _{pp}(\omega ,T)=\frac{1%
}{g\log (T/T_{\mathrm{pair}})}$. This pairing susceptibility is obviously
positive above $T_{\mathrm{pair}}$, diverges at $T_{\mathrm{pair}}$ for all $%
\omega $, and is negative below $T_{\mathrm{pair}}$, implying that the
normal state is unstable against pairing. In our case $\chi _{pp}$ can
easily be obtained from the solution of (\ref{diffeq}) by changing the
boundary condition at $x=0$ to $\Phi \left( x=0\right) =\Phi _{0}$. We then
obtain $\Phi (x)$ given by (\ref{fweak}) with 
\begin{equation}
A=\Phi _{0},~~~B=\Phi _{0}\tan (\sqrt{g}x_{T})  \label{e_10}
\end{equation}%
and 
\begin{equation}
\chi _{pp}(\omega ,T)=\frac{\cos \left( \sqrt{g}\log {\frac{\omega }{T}}%
\right) }{\cos \left( \sqrt{g}\log {\frac{\omega _{0}}{T}}\right) }.
\label{e_11}
\end{equation}%
Note that $T$ and $\omega _{0}$ are lower and upper limits of the
integration over $\omega $ in (\ref{elcomp_1}), hence the pairing
susceptibility is only defined in the interval $T<\omega <\omega _{0}$. We
see from (\ref{e_11}) that at the upper boundary, $\omega =\omega _{0}$, $%
\chi _{pp}=1$ at any $T$. This is a clear distinction to the BCS limit. As
long as $T>T_{\mathrm{pair}}$, the pairing susceptibility remains positive
everywhere in the interval $T<\omega <\omega _{0}$ despite the fact that the
solution of the differential equation (\ref{diffeq}) for $\Phi \left( \omega
\right) $ is formally an oscillating function of frequency. At $T_{\mathrm{%
pair}}$, $\log {\frac{\omega _{0}}{T}}=\frac{\pi }{2\sqrt{g}}$, and $\chi
_{pp}$ diverges for all $\omega $, except $\omega =\omega _{0}$. Below $T_{%
\mathrm{pair}}$, $\chi _{pp}$ is negative at low frequencies, implying that
the system is unstable towards pairing. We show the behavior of $\chi
_{pp}(\omega ,T)$ as function of temperature and frequency in Fig.\ref{Fig3}.

\section{Pairing problem at strong coupling}

We next analyze the strong coupling limit $g\gg 1$. In distinction to the
weak coupling regime, we now have two characteristic energy scales in the
problem, $\omega _{0}=E_{F}/g\ll E_{F}$, and $E_{F}$, which is the ultimate
upper cutoff in the theory (see Fig.\ref{Fig2}). The issue then is which of
the two scales determines the onset of the pairing.

Suppose momentarily that only frequencies $\omega \leq \omega _{0}$
contribute to the pairing. At $\omega <\omega _{0}$, the pairing kernel and
the self-energy can still be approximated by Eq. (\ref{dwwp_1}), and the
equation for the pairing vertex can still be reduced to the differential
equation (\ref{diffeq}). In distinction to the weak coupling case, however,
the term $gx$, coming from the self-energy, is now the dominant term in the
denominator in the r.h.s. of (\ref{diffeq}). Leaving only this term, we
arrive at 
\begin{equation}
\frac{d^{2}\Phi \left( x\right) }{dx^{2}}=-\frac{\Phi \left( x\right) }{x}.
\end{equation}%
Note that $g$ drops from this equation because of cancellation between $g$
factors in the effective interaction and the self-energy.

The solution of Eq.(\ref{diffeq}) with $\Phi \left( x=0\right) =0$ is $\Phi
\left( x\right) \propto \sqrt{x}J_{1}\left( 2\sqrt{x}\right) $, where $J_{1}$
is a Bessel function. Substituting this solution back into (\ref{elcomp_2})
and assuming that the upper limit in the frequency integral in (\ref%
{elcomp_2}) is still $x_{T}$ (i.e., that only $\omega <\omega _{0}$ are
relevant for the pairing), we obtain $x_{T}=3.670(5)$. This leads to $T_{%
\mathrm{pair}}\simeq 0.025\omega _{0}$, i.e., to a pairing instability at a
temperature which is a fraction of $\omega _{0}$.

This result is similar to McMillan's $T_{\mathrm{pair}}\sim \omega
_{D}e^{-(1+g)/g}\sim \omega _{D}$ for strongly coupled phonon
superconductors~\cite{McMillan68} ($\omega _{D}$ is Debye frequency).
However, like for phonons, there is actually no reason to restrict the
frequency integral to $\omega <\omega _{0}\sim E_{F}/g$, since for strong
coupling there also exists a wide frequency range $\omega _{0}<\omega <E_{F}$
where, on the one hand, the pairing kernel and the self-energy are different
from (\ref{dwwp_1}), and, on the other hand, typical frequencies are still
below $E_{F}$, i.e., a low-energy description is at least qualitatively
valid. The existence of this extra range raises the possibility that the
onset of pairing may occur at a temperature of order $E_{F}$, not of order $%
\omega _{0}\sim E_{F}/g$. Note that for the electron-phonon case, the scale
which sets the ultimate upper cutoff for the pairing (the analog of $E_{F}$
in our case) is $\omega _{D}\sqrt{g}$.\cite%
{allen,Carbotte90,Carbotte86,Combescot95}

To verify whether $T_{\mathrm{pair}}$ scales as $E_{F}$, not as $\omega _{0}$%
, we analyze the equation for $\Phi (\omega )$ assuming that all
characteristic frequencies are larger than $\omega _{0}$. At these
frequencies, the pairing kernel and the self-energy are given \ by%
\begin{eqnarray}
\Sigma \left( \omega \right) &=&-i\mathrm{sign}\left( \omega \right) g\omega
_{0}\log \frac{\left\vert \omega \right\vert }{\omega _{0}}  \nonumber \\
K\left( \omega ,\omega ^{\prime }\right) &=&\frac{\omega _{0}}{2}\left( 
\frac{1}{\left\vert \omega -\omega ^{\prime }\right\vert }+\frac{1}{%
\left\vert \omega +\omega ^{\prime }\right\vert }\right) .  \label{dwwp_2}
\end{eqnarray}%
Now the pairing kernel scales as $1/\omega $, while the self-energy is
nearly a constant, and only logarithmically depends on frequency.

Substituting the pairing kernel and the self-energy into (\ref{elcomp}) and
using the fact that for all $\omega <E_{F}$ the self-energy $\Sigma (\omega
) $ exceeds the bare $\omega $, we obtain 
\begin{equation}
\Phi \left( \omega \right) =\int_{T}^{E_{F}}\frac{\ d\omega ^{\prime }\Phi
\left( \omega ^{\prime }\right) }{\ 2\log \frac{\omega ^{\prime }}{\omega
_{0}}\ \ }\left( \frac{1}{\left\vert \omega -\omega ^{\prime }\right\vert }+%
\frac{1}{\left\vert \omega +\omega ^{\prime }\right\vert }\right) .
\label{Elfin}
\end{equation}%
The logarithmic divergence in the r.h.s. of (\ref{Elfin}) at $\omega =\omega
^{\prime }$ can easily be regularized as the $1/|\omega -\omega ^{\prime }|$
form of the kernel is only valid at $|\omega -\omega ^{\prime }|>\omega _{0}$%
.

We see that the dimensionless ratio $T/E_{F}$ is the only parameter in Eq.(%
\ref{Elfin}), except for the $\log \frac{\omega ^{\prime }}{\omega _{0}}$%
-term in the denominator in (\ref{Elfin}). Hence, if this equation has a
solution at some finite value of this parameter, the pairing instability
should occur at $T\sim E_{F}$.

The analysis of Eq. (\ref{Elfin}) requires special care because of the
interplay between the $1/\omega $ dependence of the pairing kernel and
logarithmic behavior of the self-energy. The discussion is somewhat
technical, and we moved it into Appendix A. We find there that the solution
of (\ref{Elfin}) at frequencies between $T$ and $E_F$ is 
\begin{equation}
\Phi \left( \omega \right) =A\frac{E_{F}}{\sqrt{\omega }}~\cos \left[ \beta
\log {\frac{\omega }{E_{F}}}+\phi \right] ,  \label{solution}
\end{equation}%
where $\beta =0.7923(2)$ is determined from the solution of a transcendental
equation, and $A$, $\phi $ are real constants.

Like at weak coupling, the two limits of the integration over $\omega
^{\prime }$ in Eq. (\ref{Elfin}) for $\Phi $ imply two boundary conditions
for $\Phi (\omega )$ from (\ref{solution}). One of them determines the phase 
$\phi $, while the other determines the pairing instability temperature (the
overall factor $A$ in (\ref{solution}) cannot be determined from the
linearized gap equation). For a simple estimate of $T_{\mathrm{pair}}$, we
use the same boundary conditions as in the weak coupling limit, i.e., (i)
assume that frequencies larger than $E_{F}$ are irrelevant for the pairing
and set $\Phi (\omega =E_{F})=0$, and (ii) assume that $\frac{d\Phi (\omega )%
}{d\omega }|_{\omega =T_{\mathrm{pair}}}=0$. We then obtain $\phi =\frac{\pi 
}{2}$ and 
\begin{equation}
T_{\mathrm{pair}}=E_{F}e^{-\frac{1}{\beta ^{\ast }}}\simeq 0.0676E_{F},
\label{tins}
\end{equation}%
where $1/\beta ^{\ast }=(1/\beta )\arccos \left( -\frac{1}{\sqrt{1+4\beta
^{2}}}\right) \ \simeq \frac{1}{0.371}$. In Appendix B we demonstrate that
the same result, modulo a numerical prefactor, is obtained by solving
explicitly the linearized gap equation for discrete Matsubara frequencies.

We see therefore that at strong coupling, the pairing instability
temperature $T_{\mathrm{pair}}$ is indeed of the order of the Fermi energy $%
E_{F}$, although numerically it is still much smaller than $E_{F}$. This
temperature is larger by a factor $g$ than the McMillan-type estimate, $T_{%
\mathrm{pair}}\simeq \omega _{0}$, which ignores the pairing interaction at
energies larger than the characteristic bosonic frequency. We emphasize that
in order to obtain Eq. (\ref{tins}), it was crucial that we included into
consideration the normal state self energy renormalization. Had we ignored
it, an oscillating solution for $\Phi (\omega )$ at $\omega >\omega _{0}$
would not have been possible, i.e., no pairing instability would occur at $%
T>\omega _{0}$ (see Ref.\cite{Pisarski00}). Alternatively speaking, $T_{%
\mathrm{pair}}\sim E_{F}$ is the result of the interplay between a non-Fermi
liquid behavior of the fermions caused by the logarithmic self energy $%
\Sigma \left( \omega \right) \propto g\omega _{0}\log \frac{\left\vert
\omega \right\vert }{\omega _{0}}$, and a retarded pairing interaction
governed by a \textquotedblleft local\textquotedblright\ \ boson
susceptibility $d(\omega )\propto \frac{1}{\omega }$.

Since $T_{\mathrm{pair}}\sim E_{F}$, it is inevitable that the magnitude of $%
T_{\mathrm{pair}}$ is affected by the system behavior at high energies, i.e.
at lattice scales in the condensed matter context. We assumed above that the
fermionic density of states is a constant. This is indeed only approximately
valid at $\omega \sim E_{F}$. To determine $T_{\mathrm{pair}}$ beyond an
order of magnitude estimate, one then needs to solve the full microscopic
problem. Still, lattice effects only modify the prefactor in $T_{\mathrm{pair%
}}$; the relation $T_{\mathrm{pair}}\sim E_{F}$ is generic and survives
lattice corrections.

\section{The role of gap fluctuations}

\subsection{Phase fluctuations}

In the weak coupling limit it is known that the transition temperature,
determined from the linearized gap equation, coincides with the temperature
where global phase coherency sets in. This can easily be seen by evaluating
the phase stiffness $\rho _{s}$ defined as 
\begin{equation}
E_{\mathrm{phase}}=\rho _{s}\int d^{3}x\left( \nabla \varphi \right) ^{2}.
\label{phase}
\end{equation}%
At weak coupling, $\rho _{s}\simeq E_{F}k_{F}$. Eq. (\ref{phase}) can then
be considered as the continuum limit of an $XY$- spin model on a three
dimensional lattice with lattice constant $\simeq k_{F}^{-1}$ and exchange
interaction $\simeq E_{F}$. Fluctuation effects in this model become
effective at temperatures comparable to the exchange interaction, i.e., at
temperatures comparable to the Fermi energy. Since $T_{\mathrm{pair}}\ll
E_{F}$, fluctuations at $T \sim T_{\mathrm{pair}}$ are ineffective, and
phase coherency is established as soon as Cooper pairs are formed, i.e., $T_{%
\mathrm{pair}} = T_c$.\ 

Consider next the strong coupling limit, where $T_{\mathrm{pair}}\sim E_{F}$%
. In what follows we argue that at strong coupling, $\rho _{s}/k_{F}\simeq $ 
$E_{F}/\sqrt{g}\ll T_{\mathrm{pair}}$. In this situation, phase fluctuations
become relevant well below the onset of the pairing, and by conventional
reasoning~\cite{pokrovski,emery,Franz}, phase coherence sets in at 
\begin{equation}
T_{c}\simeq E_{F}/\sqrt{g}\ll T_{\mathrm{pair}}\simeq E_{F}.  \label{t_1}
\end{equation}%
This new energy scale is the characteristic energy of a boson in the gaped
state below $T_{\mathrm{pair}}$. In between $T_{\mathrm{pair}}$ and $T_{c}$,
the systems displays a pseudogap behavior: the density of states develops a
maximum at a finite frequency (the tunneling gap), and the spectral weight
is transformed from frequencies below the gap to frequencies above the gap.
However, the superconducting order parameter only develops at $T_{c}$.

We now show how we arrived at $\rho _{s}/k_{F}\simeq $ $E_{F}/\sqrt{g}$. The
superfluid stiffness at $T=0$ is obtained by evaluating the sum of fermionic
bubbles made of normal and anomalous Green's functions, and is given by 
\begin{equation}
\rho _{s}=\rho _{s}^{0}\int_{0}^{\infty }d\omega \frac{\Phi ^{2}\left(
\omega \right) }{\left[ \left( \omega Z\left( \omega \right) \right)
^{2}+\Phi ^{2}\left( \omega \right) \right] ^{3/2}},  \label{n_11}
\end{equation}%
where $\rho _{s}^{0}\sim E_{F}k_{F}$ is the stiffness of a BCS
superconductor. $\Phi (\omega )$ is the pairing vertex at $T=0$ and we
introduced 
\begin{equation}
Z(\omega )=1-\frac{\Sigma (\omega )}{i\omega }.  \label{n_2}
\end{equation}%
Using the relation between $\Phi (\omega )$ and the gap function $\Delta
\left( \omega \right) =\frac{\Phi \left( \omega \right) }{Z\left( \omega
\right) }$, one can write Eq.(\ref{n_11}) as 
\begin{equation}
\rho _{s}=\rho _{s}^{0}\int_{0}^{\infty }\frac{d\omega }{Z(\omega )}~\frac{%
\Delta ^{2}\left( \omega \right) }{\left[ \omega ^{2}+\Delta ^{2}\left(
\omega \right) \right] ^{3/2}}.  \label{n_3}
\end{equation}%
For a BCS superconductor, $Z=1$, and $\Delta $ does not depend on frequency.
The frequency integration in (\ref{n_3}) then yields $\rho _{s}=\rho
_{s}^{0} $, independent on $\Delta $. This essentially implies that at $T=0$%
, the superfluid density equals the full density.

To obtain $\rho _{s}$ at strong coupling, we need to know $\Delta (\omega
,T=0)$ and $Z(\omega ,T=0)$. The gap $\Delta (\omega ,T=0)=\Delta (\omega )$
is obtained by solving the nonlinear gap equation 
\begin{equation}
\Delta (\omega )=\frac{3g}{2}~\int_{-\infty }^{\infty }\frac{\left( \Delta
(\omega ^{\prime })-\Delta (\omega )\frac{\omega ^{\prime }}{\omega }\right)
d_{sc}(\omega -\omega ^{\prime })}{\sqrt{\omega ^{\prime }{}^{2}+\Delta
(\omega ^{\prime })^{2}}}d\omega ^{\prime }.  \label{mats_1}
\end{equation}%
where $d_{sc}(\Omega )$ is the \textquotedblleft local\textquotedblright\
boson propagator in a superconductor. In the normal state, $d(\Omega )$ is
given by Eqs. (\ref{kernel},\ref{kernel_1}). In the presence of $\Delta $,
the bosonic spectrum itself changes due to feedback from the gap opening,
and the Landau damping transforms into $\Pi \left( \Omega \right) \sim
\gamma \frac{\Omega ^{2}}{q\Delta }\simeq \gamma \frac{\Omega ^{2}}{qE_{F}}$
(Ref.\cite{Abanov03}). This leads to 
\begin{equation}
d_{sc}\left( \Omega \right) =\frac{1}{3}\log \left( 1+\frac{\omega
_{0,sc}^{2}}{\Omega ^{2}}\right) ,  \label{t_2}
\end{equation}%
where 
\begin{equation}
\omega _{0,sc}\sim \sqrt{\omega _{0}E_{F}}\sim E_{F}/\sqrt{g}
\end{equation}%
is the characteristic energy of the bosons in a state where fermions are
gaped -- it has the same physical meaning as $\omega _{0}$ above $T_{%
\mathrm{pair}}$. For frequencies $E_{F}>|\Omega |>\omega _{0,sc}$ we have 
\begin{equation}
d_{sc}(\Omega )\simeq \frac{1}{3}\left( \frac{\omega _{0,sc}}{\Omega }%
\right) ^{2}.  \label{n_4}
\end{equation}%
Despite the $1/\Omega ^{2}$ dependence of $d_{sc}(\Omega )$, the integral
over $\omega ^{\prime }$ in Eq. (\ref{mats_1}) remains convergent since the
numerator vanishes at $\omega =\omega ^{\prime }$. We can then safely use
Eq. (\ref{n_4}) for $d_{sc}(\Omega )$ and drop the restriction that this form
is only valid above $\omega _{0,sc}$. The gap equation then contains only $%
E_{F}$ as the energy scale. Accordingly, $\Delta (\omega )$ can only be of
order $E_{F}$, if the gap equation indeed has a solution. We verified that
the solution of (\ref{n_4}) does indeed exist and yields $\Delta (\omega
)=E_{F}f(\omega /E_{F})$.

The expression for $Z(\omega )$ follows from the formula for the self-energy 
\begin{equation}
Z\left( \omega \right) =1+\frac{3\ g}{2\omega }\int_{-\infty }^{\infty }%
\frac{d\left( \omega -\omega ^{\prime }\right) \omega ^{\prime }d\omega
^{\prime }}{\sqrt{\omega ^{\prime }{}^{2}+\Delta (\omega ^{\prime })^{2}}}.
\label{n_5}
\end{equation}%
Here the restriction that Eq. (\ref{n_4}) is only valid at frequencies above 
$\omega _{0,sc}$ becomes crucial, otherwise the integral over $\omega
^{\prime }$ in Eq. (\ref{n_5}) would diverge. Beyond this, the evaluation of
the integral is straightforward, and we obtain 
\begin{equation}
Z\left( \omega <E_{F}\right) \simeq g^{1/2},~~Z\left( \omega \gg
E_{F}\right) \approx 1.  \label{n_6}
\end{equation}%
Substituting this $Z$ into (\ref{n_3}), we find 
\begin{equation}
\rho _{s}(T=0)\simeq \rho _{s}^{0}/\sqrt{g}\ \simeq \omega _{0,sc}k_{F}.
\label{n_7}
\end{equation}%
We see that $\rho _{s}(T=0)/k_{F}$ is much smaller than $T_{\mathrm{pair}}$.
\ The exchange constant of the $XY$-model (\ref{phase}) is therefore $\omega
_{0,sc}\ll E_{F}$. This leads to our estimate of $T_{c}$ in Eq. (\ref{t_1}).
This estimate is further supported by the fact that a finite $T$, we found
that the leading temperature dependence of the stiffness \ varies as a
function of $\frac{T}{\omega _{0,sc}}$, i.e., thermal corrections to the
stiffness indeed become relevant at $T$ $\simeq \omega _{0,sc}$.

\subsection{A relation to Eliashberg theory}

We emphasized above that our strong coupling theory is a local theory, but
not an Eliashberg theory. Indeed, in our case, the interaction is larger
than the Fermi energy, and the presence of the momentum cutoff in the
bosonic propagator is crucial. This distinction becomes particularly
important if we compare our result for $\rho _{s}$ with the conventional
Eliashberg theory. There $E_{F}$ is the \textit{largest} scale in the
problem, even if the \textquotedblleft local\textquotedblright\ interaction $%
d(\omega )$ scales as $1/\omega $ or even faster (as, e.g., $1/\omega ^{2}$
for phonon superconductors). Once $E_{F}$ is the largest energy scale, $\rho
_{s}/k_{F}$ is always larger than $T_{\mathrm{pair}}$, and phase
fluctuations are weak. Indeed, according to Eq. (\ref{n_11}), $\rho _{s}$
scales as 
\begin{equation}
\rho _{s}\sim \rho _{s}^{0}\frac{\Delta }{\Sigma (\omega \sim \Delta )},
\label{may29_5}
\end{equation}%
where, as before, $\rho _{s}^{0}\sim E_{F}k_{F}$ is the stiffness of the
weak coupling limit. The ratio $\Delta /\Sigma $ can be quite small if the
pairing occurs in the quantum-critical regime and involves near-massless
bosons. In particular, for phonon superconductors, when the Debye frequency $%
\omega _{D}$ is much smaller than electron-phonon interaction $u$, $\Delta
\sim T_{\mathrm{pair}}\sim u$~(see Ref.\cite{allen}), and $\Sigma (\omega
\sim u)\sim u^{2}/\omega _{D}>>\Delta $. Then $\rho _{s}\sim \rho
_{s}^{0}(\omega _{D}/u)\ll \rho _{s}^{0}$. Still, the condition that $E_{F}$
is the largest energy scale implies that $\Sigma (\omega )<E_{F}$, i.e., $%
u^{2}/\omega _{D}<E_{F}$. Then, even though $\rho _{s}$ is reduced from its
weak coupling value, it still holds that 
\begin{equation}
\rho _{s}/k_{F}\sim T_{\mathrm{pair}}\frac{E_{F}\omega _{D}}{u^{2}}>T_{%
\mathrm{pair}}.  \label{may29_5_1}
\end{equation}%
This implies that the exchange coupling in the corresponding XY model is
still larger than the onset temperature for the pairing. As a result, within
Eliashberg theory one can expect at most modest changes in the transition
temperature due to phase fluctuations. In our case, we remind, at strong
coupling $E_{F}$ is no longer the largest energy scale in the problem, and
the $1/\omega $ form of $d(\omega )$ in the strong coupling limit emerges
once one imposes a cutoff in the integration over bosonic momenta.

\subsection{longitudinal gap fluctuations}

In previous subsections we discussed the role of phase fluctuations. They
are sufficient to destroy superconducting order between $T_{c}$ and $T_{%
\mathrm{pair}}$. There also exist, however, longitudinal fluctuations of the
pairing gap, and it is instructive to consider how strong they are.

Longitudinal gap fluctuations generally reflect how shallow the profile of
the free energy with respect to deviations of $\Delta (\omega )$ from its
equilibrium value is. A shallow profile implies that the superconducting
order is weak as different $\Delta (\omega )$ have almost the same
condensation energy. A situation with a shallow profile emerges when, in
real frequencies, the attractive part of $\mathrm{Re}d_{sc}(\omega )$ is
weak, and the pairing predominantly comes from\textrm{\ }$\mathrm{Im}%
d_{sc}(\omega )$. The imaginary part of a \textquotedblleft
local\textquotedblright\ interaction describes purely retarded interaction
between fermions. This interaction then does not contribute to the
superconducting order parameter, which is an equal time correlator.
Accordingly, the slope of the free energy is determined only by a weak $%
\mathrm{Re}d_{sc}(\omega )$.

A simple estimate of the energy scale at which longitudinal gap fluctuations
become relevant can be obtained by analyzing the form of $\mathrm{Re}%
d_{sc}(\omega )$. Converting \ Eq.(\ref{t_2}) to real frequencies yields 
\begin{eqnarray}
d_{sc}\left( \Omega \right) &=&\frac{1}{3}\log \left( |1-\frac{\omega
_{0,sc}^{2}}{\Omega ^{2}}|\right)  \label{t_2_1} \\
&&+i\frac{\pi \mathrm{sign}\Omega }{3}\theta \left( \Omega ^{2}-\omega
_{0,sc}^{2}\right)  \nonumber
\end{eqnarray}%
We see that $\mathrm{Re}d_{sc}(\omega )$ remains attractive up to a
frequency $\omega _{0,sc}/\sqrt{2}$, and is repulsive at larger frequencies.
This means that frequencies above $\omega _{0,sc}/\sqrt{2}$ do not
contribute to the superconducting order parameter, although they do
contribute to the pairing itself via $\mathrm{Im}d_{sc}(\omega )$. This in
turn implies that longitudinal gap fluctuations become strong at $T\geq
\omega _{0,sc}\sim E_{F}/\sqrt{g}$. Comparing this result with Eq.(\ref{t_1}%
), we see that in our strong coupling limit, phase and amplitude
fluctuations of the gap are equally important, as the corrections to the
superconducting order parameter from both fluctuations become $O(1)$ at $%
T\sim T_{c}\sim E_{F}/\sqrt{g}$. One can equally argue that $T_{c}\ll T_{%
\mathrm{pair}}$ is the result of strong phase fluctuations, or the result of
soft longitudinal gap fluctuations brought about by the absence of a
repulsive component of $\mathrm{Re}d_{sc}(\omega )$ at $\Omega >\omega
_{0,sc}$.

\section{Migdal parameter}

As we discussed in the Introduction, the coupled equations (\ref{ELmom}) for
the pairing vertex and fermionic and bosonic self-energies are valid if
vertex corrections and the momentum dependent part of the self energy can be
neglected. In case of electron-phonon interaction, this approximation was
justified by Migdal\cite{Migal58}.Below we evaluate the leading corrections
to our local theory, both in the Eliashberg limit, and at strong coupling.
For definiteness, we focus on vertex corrections $\delta \Gamma $ of the
total vertex 
\[
\Gamma =\sqrt{\frac{u}{k_{F}}}~\left( 1+\delta \Gamma \right) . 
\]

\begin{figure}[tbp]
\label{Fig4} \includegraphics[width=6.0cm]{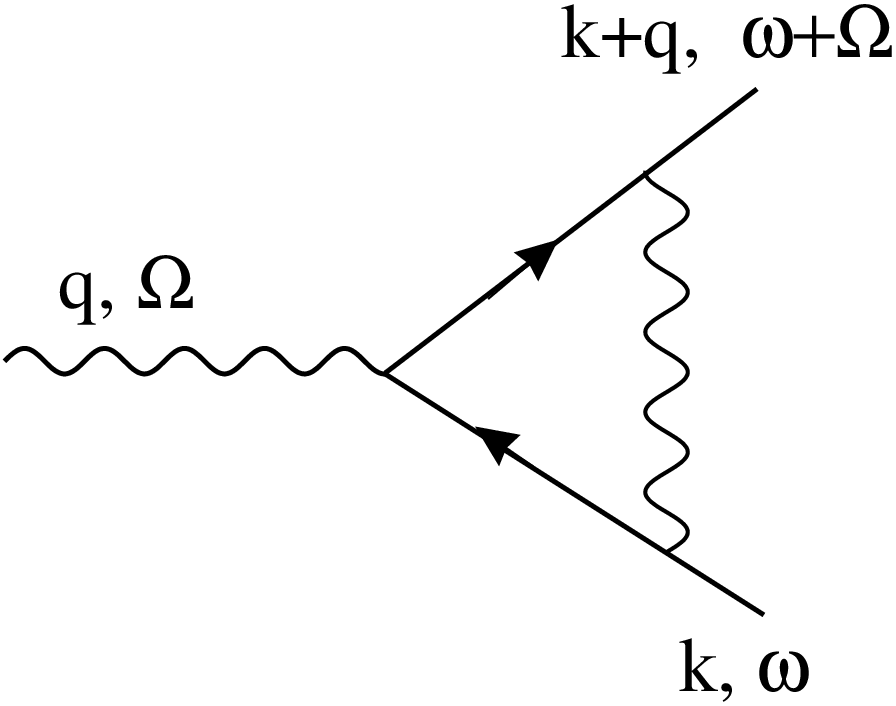}
\caption{Leading correction of the fermion boson vertex $\Gamma _{k,q}\left( 
\protect\omega ,\Omega \right) $. The solid lines stand for the fermionic \
propagator and the wiggly lines for the bosonic propagator, respectively.}
\end{figure}

Generally, the correction to the interaction vertex between fermions and
gapless bosons, depend on the interplay between the bosonic momentum and
frequency. In particular, Ward identities imply that vertex corrections in
the limit of vanishing bosonic momentum are of the same order as the
fermionic self-energy, and not necessary small\cite{Chubukov04}. However,
for the pairing problem, we need to analyze vertex corrections for typical
bosonic energies, $\Omega _{\mathrm{typ}}$, and for typical bosonic momenta, 
$q_{\mathrm{typ}}$, that contribute to the pairing.

The leading vertex correction in the normal state is presented in Fig.\ref%
{Fig4} and is given by:%
\begin{eqnarray}
\delta \Gamma _{q}\left( \omega ,\Omega \right) &=&\frac{Nu}{k_{F}}%
\int_{k^{\prime }\omega ^{\prime }}D_{k-k^{\prime }}\left( \omega -\omega
^{\prime }\right) G_{k^{\prime }}\left( \omega ^{\prime }\right)  \nonumber
\\
&&\times G_{k^{\prime }+q}\left( \omega ^{\prime }+\Omega \right) .
\label{t_3}
\end{eqnarray}%
Here, $\omega $ is the external fermionic frequency. In principle, $\delta
\Gamma $ depends on two momenta -- the bosonic momentum $q$ and \ the
external fermionic momentum $k$. However, the dependence on $k$ is weak and
thus irrelevant, and we will neglect it. Performing the momentum integration
in (\ref{t_3}), we obtain%
\begin{equation}
\delta \Gamma _{q}\left( \omega ,\Omega \right) \simeq \frac{%
3g\int_{0}^{\Omega }d\left( \omega +\omega ^{\prime }\right) d\omega
^{\prime }}{\sqrt{\Omega ^{2}+\left( Nv_{F}q\right) ^{2}}}.  \label{vertex}
\end{equation}%
The factor $N$ in the denominator is a consequence of the rescaling that we
performed in Sec. II. In the limit $q\rightarrow 0$ 
\begin{eqnarray}
\delta \Gamma _{q\rightarrow 0}\left( \omega ,\Omega \right) &=&\frac{3g}{%
\Omega }~\int_{\omega }^{\Omega +\omega }d\left( \omega ^{\prime }\right)
d\omega ^{\prime }  \label{vertex_1} \\
&=&\frac{\Sigma \left( \Omega +\omega \right) -\Sigma \left( \omega \right) 
}{-i\Omega }.  \nonumber
\end{eqnarray}%
This is the Ward identity relating the homogeneous vertex with the self
energy. We see from (\ref{vertex_1}) that static vertex corrections do not
depend on $N$ and are not small at moderate and strong coupling. The
situation, however, changes when we evaluate $\delta \Gamma _{q}\left(
\omega ,\Omega \right) $ at $q_{\mathrm{typ}}$ and $\Omega _{\mathrm{typ}}$
relevant to the pairing problem. In what follows we evaluate $\delta \Gamma $
for weak, intermediate and strong coupling, and specify in each case the
relevant bosonic momentum and frequency.

\subsection{Vertex corrections at weak coupling}

We first consider the limit of weak coupling, $g\ll 1$. The typical bosonic
energy $\Omega _{\mathrm{typ}}$ is of order $T_{c}\simeq (E_{F}/g)e^{-\pi /(2%
\sqrt{g})}$. On the other hand, typical momenta $q_{\mathrm{typ}}$ are
obtained from the condition that the momentum and frequency dependent term
in the bosonic propagator $D(q,\Omega )$ are of the same order, i.e. $q_{%
\mathrm{typ}}\simeq \left( \gamma \Omega _{\mathrm{typ}}\right) ^{1/3}$.
Together with Eq.(\ref{e_5_1}) for $\gamma $ this yields 
\begin{equation}
v_{F}q_{\mathrm{typ}}\simeq E_{F}e^{-\pi /(6\sqrt{g})}.  \label{m_1}
\end{equation}%
Comparing $\Omega $ and $v_{F}q$, we see that 
\begin{equation}
v_{F}q_{\mathrm{typ}}\simeq \Omega _{\mathrm{typ}}\left[ ge^{\pi /(3\sqrt{g}%
)}\right] \gg \Omega _{\mathrm{typ}}.  \label{m_2}
\end{equation}%
We then obtain from Eq.(\ref{vertex}) 
\begin{equation}
\delta \Gamma \simeq \frac{g\int_{0}^{\Omega _{\mathrm{typ}}}\log \left( 
\frac{\omega _{0}}{\omega }\right) d\omega ^{\prime }}{Nv_{F}q_{\mathrm{typ}}%
}=\frac{\pi }{2N}\frac{e^{-\frac{\pi }{3\sqrt{g}}}}{\sqrt{g}}.
\end{equation}%
We see that vertex correction is exponentially small for small $g$. The
extension to large $N$ is in fact not needed as vertex corrections are
already negligible.

\subsection{Vertex corrections at intermediate coupling}

At intermediate $g=O(1)$, $\omega _{0}$ and $E_{F}$ become of the same
order, i.e., there is only one characteristic energy scale in the problem.
Vertex corrections are $O(1)$ for $N=1$, but are still small in $1/N$ if we
extend the theory to large $N$.

\subsection{Vertex corrections at strong coupling}

A naive expectation would be that vertex corrections gradually increase with 
$g$ and eventually overcome the overall smallness in $1/N$. \ This would
invalidate our local theory at sufficiently large $g$. It turns out,
however, that vertex corrections freeze at $O(1/N)$ and do not grow with $g$%
. The saturation originates from the form of the bosonic propagator $%
D(q,\Omega )$, which \ is determined by the self energy $\simeq \gamma \frac{%
|\Omega |}{q}$. As $\gamma $ by itself scales with the boson-fermion
coupling, $D(q,\omega )$ scales inversely with $g$ and cancels out the
overall factor $g$ in Eq.\ref{vertex}.

To see this explicitly we note that at strong coupling, $q_{\mathrm{typ}}$
is of order $k_{F}$, hence $v_{F}q_{\mathrm{typ}}$ is of order $E_{F}$.
Typical frequencies $\Omega _{\mathrm{typ}}$ are also $O(E_{F})$. The
external fermionic frequency $\omega $ in Eq. (\ref{vertex}) is also of
order of the Fermi energy. Evaluating the vertex correction diagram using $%
d\left( \Omega \right) \simeq \frac{\omega _{0}}{3\left\vert \Omega
\right\vert }$, we obtain for these $q_{\mathrm{typ}}$ and $\Omega _{\mathrm{%
typ}}$ 
\begin{equation}
\delta \Gamma \simeq \frac{g\int_{0}^{E_{F}}\frac{\omega _{0}}{E_{F}+\omega
^{\prime }}d\omega ^{\prime }}{E_{F}\sqrt{1+N^{2}}}\simeq \frac{1}{N}
\label{m_4}
\end{equation}%
We see that vertex corrections indeed do not depend on $g$ and remain small
(at large $N$) for arbitrary strong coupling.

At large $N$, vertex corrections remain small for all values of $g$, i.e.,
our local theory is valid both in the weak and the strong coupling limit.

\section{Other systems}

As we discussed in the introduction, the problem discussed in this paper
describes $p-$wave superconductivity in condensed matter systems close to a
ferromagnetic quantum critical point with Ising symmetry\cite{comm_3}, and
superconductivity due to color magnetic interactions in dense quark matter%
\cite{Son99,Schaefer99,Pisarski00}. As pointed out above, for color
superconductivity, antiparticle pairing that has been neglected in Eqs. (\ref%
{ELmom}), (\ref{ELmom_1}), may come into play at strong coupling\cite%
{Pisarski02}. We nevertheless believe that the main results of this paper
are still relevant to this case.

However, the results obtained in the strong coupling limit are much more
general. The key aspect of our theory is that the boson propagator $D(%
\mathbf{q}_{\parallel },q_{\perp },\Omega )$ becomes completely local above
a certain frequency. We considered a particular case when boson dynamics is
set by Landau damping, and 
\begin{equation}
D\left( \Omega \right) \propto \frac{\omega _{0}}{\left\vert \Omega
\right\vert },  \label{dgen}
\end{equation}%
where $\omega _{0}$ is a characteristic upper cut off scale of the boson
system. However, our results will be equally valid for any $D(\Omega )$ in
the form $D(\Omega )\propto 1/\left\vert \Omega \right\vert ^{\gamma }$ with 
$\gamma \geq 1$.

One of the possible applications of our strong coupling result is pairing by
antifereromagnetic fluctuations, which has been discussed in great detail in
the context of cuprate superconductors\cite{Abanov03}. That analysis was,
however, performed within the low-energy spin-fermion model, which is only
valid at $u<E_{F}$, i.e., in the weak coupling limit, as we defined it in
this paper. In the cuprates, as is well known, the Hubbard interaction is
comparable to $E_{F}$, i.e., $g\geq 1$. Our finding that a large pseudogap
regime with phase incoherent pairs is inevitable at strong coupling is quite
intriguing in view of the numerous observations of pseudogap physics in this
class of materials. To make a connection to other studies of pseudogap
within Hubbard model, we note that our effective interaction $u\simeq
U^{2}/t $ and $E_{F}\simeq t$, where $U$ and $t$ are the local Coulomb
repulsion and tight binding hopping element, respectively. Then $g=\left( 
\frac{U}{t}\right) ^{2}$ and $\omega _{0}\sim E_{F}/g\propto \frac{t}{U}J$,
and $\omega _{0,s}\sim E_{F}/\sqrt{g}\sim J$, where $J=\frac{4t^{2}}{U}$ is
the antiferromagnetic exchange interaction between spins. Accordingly, it
follows from our analysis that the pairing sets in at $T_{\mathrm{pair}}\sim
t$, and the pairing gap is $\Delta \sim t$, while coherent superconductivity
occurs at $T_{c}\propto J$. We note in this regard that numerical
investigations of variational wave functions designed to cover the strong
coupling limit\cite{Anderson87} do indeed yield a zero temperature gap $%
\Delta \sim t$ and a considerably reduced superfluid stiffness $\rho _{s}$%
\cite{Gros87,Paramekanti01,Liu05}. Also, our result $T_{c}\sim J$ is
consistent with the observation that $T_{c}$ in underdoped cuprates scales
with the neutron peak frequency. The latter turns out to be of order $J$,
if, e.g., one extends the exciton scenario for the resonance peak to the
strong coupling limit~\cite{comm_4}.

There exists some similarity between our results and those obtained for the
crossover from BCS-type behavior at weak coupling and Bose Einstein
condensation (BEC) of pairs at strong 
coupling\cite{Legett80,Nozieres85,Scalettar89,Melo93}.
In particular, for large $g$, when $\Delta 
$ is or order $E_{F}$, our pair coherence length $\xi \simeq
v_{F}/\Delta $ becomes of the order of the typical distance between fermions
($\sim k_{F}^{-1}$). This is similar to the findings for the BCS-BEC
crossover\cite{Legett80,Nozieres85,Scalettar89,Melo93}. An important
distinction between the two theories is that in BCS-BEC crossover,the
pairing interaction is static, while in our case it is dynamic and strongly
retarded. The transition at large $g$ in our case 
should therefore not be considered as
condensation of almost free bosons.

\section{Conclusion}

In this paper we considered pairing of 3D fermions, due to an exchange of
massless bosons. The model we considered describes $p-$wave
superconductivity in itinerant fermionic systems close to a ferromagnetic
quantum critical point with Ising symmetry, and superconductivity due to
color magnetic interactions in quark matter.

At weak coupling, we find that an exchange of a massless boson enhances $%
T_{c}$ compared to the BCS expectation $T_{c}^{\mathrm{BCS}}\propto e^{-%
\frac{1}{g}}$, and the actual transition temperature is $T_{c}\simeq \omega
_{0}e^{-\frac{\pi }{2\sqrt{g}}}$, where $\omega _{0}=E_{F}/g$ is a
characteristic energy scale of the bosons. This result agrees with previous
calculations\cite{Son99,Schaefer99,Pisarski00}.

At strong coupling, we find that pairing emerges at a temperature $T_{%
\mathrm{pair}}\simeq 0.06E_{F}$, which is only numerically smaller than the
Fermi energy. In addition, we find that the phase stiffness behaves as $\rho
_{s}\sim T_{\mathrm{pair}}/\sqrt{g}$, i.e. the typical energy scale where
phase fluctuation become important is parametrically smaller than $T_{%
\mathrm{pair}}$. The small phase stiffness implies that coherent
superconductivity only emerges at $T_{c}\simeq T_{\mathrm{pair}}/\sqrt{g}\ll
T_{\mathrm{pair}}$. In between $T_{c}$ and $T_{\mathrm{pair}}$ the system
displays pseudogap behavior with preformed pairs. The width of the pseudogap
regime widens as $g$ grows.

We further argued that several aspects of the strong coupling solution,
particularly the existence of two temperatures $T_{\mathrm{pair}}$ and $%
T_{c} $, are valid for a much larger class of problems in which boson
propagator becomes completely local above a certain frequency. \ The
existence of a cut off energy scale, which corresponds to the finite lattice
constant in the condensed matter context or to the large density in case of
color superconductivity, is crucial for the relevance of phase fluctuations.
For the cuprates, our results imply that $T_{\mathrm{pair}}$ and the pairing
gap $\Delta $ scale with the hopping integral $t$, while $T_{c}$ scales with
the exchange interaction $J$.

\begin{acknowledgments}
\ This work was supported in part by the National Science Foundation grants
NSF DMR02-40238 (A.C.) and PHY99-07949 and by the Ames Laboratory, operated
for the U.S. Department of Energy by Iowa State University under Contract
No. W-7405-Eng-82 (J.S.). Part of this work was carried out when both of us
participated in the 2005 workshop on "Quantum phase transitions" at the
Kavli Institute for Theoretical Physics. We acknowledge useful 
 conversations with M. Dzero, P. Coleman, V. Khodel, D. Maslov, C. Pepin, and 
 V. Yakovenko.
\end{acknowledgments}

\bigskip\ 

\appendix

\section{Pairing with $\frac{1}{\left\vert \protect\omega -\protect\omega %
^{\prime }\right\vert }$ kernel}

In this appendix we obtain the solution of the Eliashberg equation, Eq.(\ref%
{Elfin}), in the strong coupling regime. The analysis of a pairing problem
with a kernel $\frac{1}{\left\vert \omega -\omega ^{\prime }\right\vert }$
is nontrivial and it is useful to \ solve a more general problem first~\cite{Altshuler05}. We
consider:

\begin{equation}
h_{\gamma }\left( \omega \right) \ =\frac{1-\gamma }{2}\int_{0}^{\infty }%
\frac{\ d\omega ^{\prime }h_{\gamma }\left( \omega ^{\prime }\right)
k_{\gamma }\left( \omega ,\omega ^{\prime }\right) }{\ (\omega
^{\prime})^{1-\gamma }\ }  \label{hw1}
\end{equation}%
with%
\begin{equation}
k_{\gamma }\left( \omega ,\omega ^{\prime }\right) =\frac{1}{\left\vert
\omega -\omega ^{\prime }\right\vert ^{\gamma }}+\frac{1}{\left\vert \omega
+\omega ^{\prime }\right\vert ^{\gamma }}
\end{equation}%
and argue below that $\Phi \left( \omega \right) \simeq \lim_{\gamma
\rightarrow 1}h_{\gamma }\left( \omega \right) $.

The integral equation (\ref{hw1}) is scale invariant suggesting a power-law
solution 
\begin{equation}
h_{\gamma }\left( \omega \right) =A\omega ^{-b}.
\end{equation}%
Inserting this ansatz into Eq. (\ref{hw1}) we obtain 
\begin{equation}
1=\frac{1-\gamma }{2}\ \int_{0}^{\infty }dt\frac{1}{t^{1+b-\gamma }}\left( 
\frac{1}{\left\vert t-1\right\vert ^{\gamma }}+\frac{1}{\left\vert
t+1\right\vert ^{\gamma }}\right) ,  \label{ch_1}
\end{equation}%
where $t=\frac{\omega ^{\prime }}{\omega }$. This determines the exponent $%
b\left( \gamma \right) $. The integral over $t$ can be performed explicitly.
In the limit where $\gamma $ is close to $1$, Eq. (\ref{ch_1}) reduces to 
\begin{equation}
1=\ 1+\left( 1-\gamma \right) y\left( b\right) ,  \label{ch_2}
\end{equation}%
where $y\left( b\right) =\gamma _{E}+\psi \left( b\right) -\frac{\pi }{2}%
\tan \left( \frac{b\pi }{2}\right) $, and $\psi \left( b\right) $ is the
di-gamma function. While $b$ is undetermined for $\gamma =1$ it must hold
that $y\left( b\right) =0$ for any $\gamma \neq 1$. For real $b$ the
condition $y\left( b\right) =0$ cannot be fulfilled. However, for a complex $%
b=\alpha +i\beta $, we find that the imaginary part of $y\left( b\right) $
vanishes if $\alpha =\frac{1}{2}$, i.e. if $y\left( \frac{1}{2}+i\beta
\right) $ is purely real. Using this fact and substituting $b=1/2+i\beta $
into (\ref{ch_2}), we obtain that $\beta $ is determined from 
\begin{equation}
Re\left( \gamma _{E}+\psi \left( \frac{1}{2}+i\beta \right) \right) =\frac{%
\pi }{2}\tan \left( \frac{\pi }{4}+i\frac{\pi \beta }{2}\right) .
\end{equation}%
This equation is easily solved graphically and yields $\beta =\pm 0.7923(2)$%
. Therefore 
\begin{equation}
h_{\gamma }\left( \omega \right) =A\mathrm{\ }\omega ^{-\frac{1}{2}-i\beta
}+A^{\ast }\omega ^{-\frac{1}{2}+i\beta }.  \label{sol}
\end{equation}%
The overall constant $A$ is chosen such that $h_{\gamma }\left( \omega
\right) $ is real.

In order to show that Eq.(\ref{sol}) is indeed the solution of Eq.(\ref%
{Elfin}) we discuss more carefully why $\lim_{\gamma \rightarrow 1}h_{\gamma
}\left( \omega \right) $ gives the desired $\Phi \left( \omega \right) $.
Eq. (\ref{Elfin}) can be re-expressed as 
\begin{equation}
\Phi \left( \omega \right) =\lim_{\gamma \rightarrow 1}\frac{1-\gamma }{2}%
\int_{T}^{E_{F}}\frac{\ d\omega ^{\prime }\Phi \left( \omega ^{\prime
}\right) k_{\gamma }\left( \omega ,\omega ^{\prime }\right) }{\ \omega
^{\prime 1-\gamma }-\omega _{0}^{1-\gamma }\ }.\   \label{ch_3}
\end{equation}%
This equation coincides with (\ref{hw1}) if we neglect $\omega
_{0}^{1-\gamma }$ in the denominator. We now recall that at strong coupling, 
$T_{\mathrm{pair}}\gg \omega _{0}$. Then for all $\omega ^{\prime }>T$ in (%
\ref{ch_3})holds that $\omega ^{\prime }\gg \omega _{0}$, and we can safely
neglect $\omega _{0}^{1-\gamma }$ for any $\gamma \neq 1$.

\section{Linearized gap equation for discrete Matsubara frequencies}

In the computation of $T_{\mathrm{pair}}$ in the main text we imposed the
lower cutoff in the zero-temperature equation for the pairing vertex. In the
weak coupling limit, this procedure was shown earlier~\cite{Pisarski00} to
yield the same $T_{\mathrm{pair}}$ (modulo a numerical prefactor), as one
would obtain by performing an explicit summation over discrete Matsubara
frequencies. \ In this appendix we demonstrate that the same is true in the
strong coupling limit.

The most straightforward way to analyze the linearized pairing problem is by
considering the gap function%
\begin{equation}
\Delta _{n}=\frac{\Phi \left( \omega _{n}\right) }{Z\left( \omega
_{n}\right) },
\end{equation}%
where $Z\left( \omega _{n}\right) =1-\frac{\Sigma \left( \omega _{n}\right) 
}{i\omega _{n}}$ and determine the temperature at which $\Delta _{n}\neq 0$
for the first time. The advantage of analyzing $\Delta $ instead of $\Phi $
is that the corresponding linearized gap equation does not explicitly
contain the fermionic self-energy and can more easily be solved numerically.
The equation for $\Delta _{n}$ is straightforwardly obtained from the
equations for $\Phi \left( \omega _{n}\right) $ and $\Sigma \left( \omega
_{n}\right) $:%
\begin{eqnarray}
\Delta _{n} &=&3\pi gT~\sum_{m}\left( \frac{\Delta _{m}}{\omega _{m}}-\frac{%
\Delta _{n}}{\omega _{n}}\right)  \nonumber \\
&&\times \mathrm{sign}\left( \omega _{m}\right) d\left( \omega _{m}-\omega
_{n}\right) .  \label{matsgap}
\end{eqnarray}%
In the limit where $T_{\mathrm{pair}}>\omega _{0}$, relevant $|\omega
_{n}|>\omega _{0}$, we can approximate $d(\Omega _{n})$ by $d\left( \Omega
_{n}\right) \simeq \frac{\omega _{0}}{3\left\vert \Omega _{n}\right\vert }$.
Then 
\begin{eqnarray}
\Delta _{n} &=&\frac{E_{F}}{4\pi T_{\mathrm{pair}}}~\sum_{m\neq n}\left( 
\frac{\Delta _{m}}{m+1/2}-\frac{\Delta _{n}}{n+1/2}\right)  \nonumber \\
&&\times \frac{\mathrm{sign}\left( m+1/2\right) }{\left\vert m-n\right\vert }%
.  \label{mathe2}
\end{eqnarray}%
The summation over discrete Matsubara frequencies is convergent for large $n$%
. Thus, no regularization or cut off is needed to solve for $\Delta _{m}$.
The only dimensionless parameter in (\ref{mathe2}) is $E_{F}/T_{\mathrm{pair}%
}$, and the non-trivial solution of (\ref{mathe2}), if it exists, appears at 
$T_{\mathrm{pair}}\simeq E_{F}$ as we found earlier. We verified numerically
that the solution of Eq. (\ref{mathe2}) does indeed exist at $T_{\mathrm{pair%
}}\simeq 0.064E_{F}$. This is even quantitatively close to the estimate $T_{%
\mathrm{pair}}\simeq 0.0676E_{F}$ obtained in the main text.


\begin{thebibliography}{99}
\bibitem{comm_1} The condition $g\ll 1$ near SDW or CDW instabilities can
still be valid in some cases, but requires extra assumptions about the
behavior of the interaction $u(r)$ in real space ~\cite{Dzero04}.
\bibitem{comm_2} Physically, a finite $\Phi _{0}$ can be realized by
coupling our system to another superconductor through a weak Josephson
junction.
\bibitem{comm_3} The restriction to Ising systems is important as the
transition becomes otherwise of first order~~\cite{Chubukov03}.
\bibitem{comm_4} In the exciton scenario, the resonance peak frequency is $%
\Omega _{res}\sim (\Delta \omega _{sf})^{1/2}$ where $\Delta $ is a gap, and 
$\omega _{sf}$ is the spin-fluctuation frequency. At weak coupling, $\Delta
\sim {\bar{g}}$, and $\omega _{sf}\sim v_{F}^{2}\xi ^{-2}/{\bar{g}}$, where $%
v_{F}\sim t$ is Fermi velocity, ${\bar{g}}\sim u^{2}/t$, and $\xi $ is a
magnetic correlation length. Accordingly, $\Omega _{res}\sim v_{F}\xi ^{-1}$%
. At strong coupling, $\Delta \sim t$, $\omega _{sf}\sim t^{3}/u^{2}$, and $%
\Omega _{res}\sim J\xi ^{-1}$.
\end{thebibliography}
\end{document}